\documentclass[twocolumn,epjc3]{svjour3}

\RequirePackage{fix-cm}

\smartqed

\RequirePackage{graphicx}
\usepackage{caption}
\usepackage{subcaption}
\RequirePackage{mathptmx}
\RequirePackage{flushend}
\RequirePackage{latexsym}
\RequirePackage[numbers,sort&compress]{natbib}
\RequirePackage[colorlinks,citecolor=blue,urlcolor=blue,linkcolor=blue]{hyperref}

\journalname{Eur. Phys. J. C}

\begin{document}

\title{Lie symmetry approach to the time-dependent Karmarkar condition}

\author{Andronikos Paliathanasis \thanksref{addr1,e1}
       \and
             Robert S. Bogadi \thanksref{addr2,e2}
        \and
              Megandhren Govender \thanksref{addr2,e3}
}

\thankstext{e1}{e-mail: anpaliat@phys.uoa.gr}
\thankstext{e2}{e-mail: bogadi.robert@gmail.com}
\thankstext{e3}{e-mail: megandhreng@dut.ac.za}

\institute{Institute of Systems Science, Durban University of Technology, Durban 4000, South Africa \label{addr1}
           \and
               Department of Mathematics, Faculty of Applied Sciences, Durban University of Technology, Durban 4000, South Africa \label{addr2}
                              }

\date{Received: date / Accepted: date}

\maketitle

\begin{abstract}
We obtain solutions of the time-dependent Einstein Field Equations which satisfy the Karmarkar condition via the method of Lie symmetries. Spherically symmetric spacetime metrics are used with metric functions set to impose conformal flatness, Weyl-free collapse and shear-free collapse. In particular, a solution was found which satisfies the heat-flux boundary condition of Santos, and a radiating stellar model was then obtained and investigated. Solutions obtained which do not allow for the application of the junction conditions at a boundary surface may lend themselves to cosmological models. This is a first attempt in generating solutions satisfying the Karmarkar condition via the method of Lie symmetries and our example of a radiating model highlights the viability of this method.
\end{abstract}


\maketitle


\section{Introduction}

The gravitational collapse of stellar objects is of much interest in relativistic astrophysics, requiring the solution of time-dependent relativistic field equations. Gravitational collapse problems were pioneered by Oppenheimer and Synder \cite{opp}, not long after General Relativity was formulated by Einstein. Initially, the Schwarzschild solution was used until the discovery of the Vaidya solution \cite{vaid} which accommodates null-radiation in the exterior atmosphere due to energy being radiated away from the collapsing body of fluid. There have been numerous attempts at obtaining solutions of the Einstein field equations for describing a radiating body, simultaneously undergoing gravitational collapse, and these efforts typically employ boundary conditions, equations of state, initial static configurations and separation of variables \cite{bon,pret,bog}. The boundary of a collapsing star divides the spacetime into two distinct regions, the interior, $\mathcal{M}^-$, and the exterior spacetime, $\mathcal{M}^+$. The interior spacetime must match smoothly to the exterior spacetime in order to generate a physically viable model of a radiating star. Early attempts were made by Glass \cite{glas} in which the Darmois and Lincherowitz matching conditions were utilised. Santos then established the appropriate boundary conditions for a spherically symmetric, shear-free, time-dependent metric which matches smoothly to the exterior Vaidya metric \cite{san}. The Santos matching condition requires a non-vanishing pressure at the boundary for a star dissipating energy in the form of a radial heat flux. This is a necessary condition that ensures continuity of momentum flux across the boundary. The Santos junction conditions have been generalised to include shear \cite{nai,thir1}, the cosmological constant as well as the electromagnetic field \cite{mah1,thir2}. Herrera and co-workers \cite{her1,her2,her3} have established important, fundamental results concerning matter distribution, stability of the shear-free condition, energy conditions and thermodynamic properties of gravitational collapse processes. \\
In developing models describing gravitational collapse, assumptions concerning the gravitational potentials and matter content of the gravitating body are often made. These have included acceleration-free and expansion-free collapse, Weyl-free collapse, anisotropic pressure configurations, the inclusion of shear and bulk viscosity and stipulations of equation of state \cite{mah2,gov,red}. Differential equations which arise, typically with respect to invariance of the junction conditions, lend themselves to the application of Lie symmetry methods and this can help to determine novel, exact solutions  \cite{mso,abe1,abe2}. In addition to restricting the matter content, conditions on the spacetime geometry are just as important. It is of interest to consider gravitational fields, typically represented by a Riemannian metric of four dimensions, to be immersed in a flat spacetime of higher dimension. Randall-Sundram and Anchordoqui-Bergliaffa re-established the conjecture that 4-dimensional spacetime might be embedded in higher dimensional flat space and much effort is made to achieve class one embedding \cite{mau1}. In general, an n-dimensional Riemannian spacetime is said to be of class p if it can be embedded into a flat space of dimension n + p. The Karmarkar condition \cite{kar} is a necessary (but not sufficient) condition for a spacetime to be of class one \cite{pan}. The derivation of the Karmarkar condition is purely geometric in nature providing relations among the components of the Riemann tensor. This in turn relates the metric potentials to one another and has thus assisted in model development. Of interest is to note that the Karmarkar condition together with pressure isotropy gives the interior Schwarzschild solution as the only bounded matter configuration. Recent attempts at modelling compact objects such as 4U 1538-52,PSR J1614-2230, Vela X-1 and Cen X-3 using the Karmarkar condition have produced models with favourable physical characteristics, consistent with observations \cite{bha,mau2}. It is expected that the Karmarkar condition should also be favourable for time-dependent systems, perhaps improving stability which is an issue for shear-free spacetimes \cite{her2}.\\
Lie symmetry analysis is a very powerful tool for the study of nonlinear differential equations and we make use of this in generating solutions of differential equations obtained via Karmarkar's condition. The steps that we follow in the analysis are: (i) we determine the Lie symmetries for each master partial differential equation for each model, (ii) the one-dimensional optimal system is determined in each case for the admitted Lie symmetries, (iii) we define similarity transformations from the Lie symmetries by using the Lie invariants which are used to reduce the master equation into an ordinary differential equation and (iv) exact closed-form solutions are then obtained. These steps have been applied before for various
gravitational models with interesting results. Exact solutions describing charged radiation have been derived by applying Lie symmetries \cite{abe4}. Moreover, in \cite{abe2} Lie symmetries have been used to derive expanding and shearing models of radiating relativistic stars, while shear-free radiating stars were considered in \cite{abe3}. In \cite{GG}, a new solution which describes a Euclidean star is derived by Lie symmetries and has the physical property of satisfying all the energy conditions and admitting a barotropic equation of state. For the field equations of the Schwarzschild model, conservation laws, invariant functions and differential operators derived by using the Lie symmetry analysis in \cite{ch5,ch6}. The Emden--Fowler equation which can describe gravitational spherically symmetric solutions was investigated via symmetry analysis in \cite{ch7}. For other applications of symmetry analysis in
gravitational physics, we refer the reader to \cite{ch8,ch9,ch10,ch11} and the references therein. \\
The plan of this paper is as follows: In Section 2 we present the gravitational field equations for our analysis. In Section 3, the basic properties and definitions of Lie symmetries are given. The definition of the one-dimensional optimal system is discussed. The latter is necessary in order to perform a complete classification of the admitted similarity transformations. In Sections 4, 5 and 6 we solve the Karmarkar condition with respect to conformally flat, Weyl-free collapse and shear-free metrics respectively. A physical application is then given in Section 7 which describes radiating, Weyl-free collapse. In Section 8 we discuss the merits of the radiating stellar model obtained and in Section 9 we conclude on the suitability and novelty of our methods. Appendices A, B and C complete the presentation of this study where we present exact solutions for the three cases given in Sections 4, 5 and 6. These may be used for future studies.

\section{Spherically symmetric spacetimes in relativity}

The spherically symmetric line element is given by

\begin{equation}
ds^2 = -A^2 \left(t, r\right) dt^2 + B^{2}\left(t, r\right)  dr^2 + Y^2 \left(t, r\right)  d\Omega^2 \label{le.01}
\end{equation}

where$~d\Omega^2$ is the line element of the two-sphere, that is,

\begin{equation}
d\Omega^2 = d\theta^2 + \sin^2\theta d\phi^2. \label{le.02}
\end{equation}

An energy-momentum tensor incorporating heat flux, $q^{a} = (0,q^1,0,0)$, is used,

\begin{eqnarray} \label{energymom}
T^-_{ab} &=& (\rho + p_{t})u_au_b + p_{t}g_{ab}+ (p_{r}-p_{t}) \chi_{a} \chi_{b} + q_au_b + q_bu_a  \nonumber \\  
\end{eqnarray}

where, $\rho$ is the energy density, $p_{r}$ the radial pressure, $p_{t}$ the tangential pressure and $q_a$ the heat flux vector. The timelike four-velocity of the fluid is $u_a$ and $\chi_a$ is a spacelike unit four-velocity along the radial direction. These quantities must satisfy $u_a u^a = -1$, $u_a q^a = 0$, $\chi_a \chi^a = 1$ and $\chi_a u^a = 0$. In co-moving coordinates we have

\begin{equation}
 u^a = A^{-1} \delta_{0}^{a} ~~,~~  q^a = q \delta_{1}^{a} ~~,~~  \chi^a = B^{-1} \delta_{1}^{a}
 \end{equation}
 
 The four-acceleration and expansion scalar are given by
 
 \begin{equation}
 w_a = u_{a;b} u^b  ~~,~~  \Theta = {u^a}_{;a}
 \end{equation}

Einstein's time-dependent field equations are then given by

\begin{eqnarray}
	\rho &=& \frac{1}{A^2}\left(2\frac{\dot B}{B} + \frac{\dot Y}{Y}\right)\frac{\dot Y}{Y}  \nonumber \\
	&& - \frac{1}{B^2}\left[2\frac{Y''}{Y} + \left(\frac{Y'}{Y}\right)^2 - 2\frac{B'}{B}\frac{Y'}{Y} -\left(\frac{B}{Y}\right)^2\right], \label{EFE1AA}
\end{eqnarray}
\begin{eqnarray}
	p_r &=& -\frac{1}{A^2} \left[2\frac{{\ddot Y}}{Y} - \left(2\frac{\dot A}{A} - \frac{\dot Y}{Y}\right)\frac{\dot Y}{Y}\right] \label{EFE2AA} \nonumber \\ 
	&& + \frac{1}{B^2}\left(2\frac{A'}{A} + \frac{Y'}{Y}\right)\frac{Y'}{Y}- \frac{1}{Y^2},
\end{eqnarray}
\begin{eqnarray}
	p_{t} &=& -\frac{1}{A^2}\left[\frac{\ddot B}{B} + \frac{{\ddot Y}}{Y} - \frac{\dot A}{A}\left(\frac{\dot B}{B} + \frac{\dot Y}{Y}\right) + \frac{\dot B}{B}\frac{\dot Y}{Y}\right]  \nonumber \\&&+ \frac{1}{B^2}\left[\frac{A''}{A} + \frac{Y''}{Y} - \frac{A'}{A}\frac{B'}{B} + \left(\frac{A'}{A} - \frac{B'}{B}\right)\frac{Y'}{Y}\right], \label{EFE3AA}
\end{eqnarray}
\begin{eqnarray}
	q &=& \frac{2}{AB}\left(\frac{\dot{Y'}}{Y} - \frac{\dot B}{B}\frac{Y'}{Y} - \frac{\dot Y}{Y}\frac{A'}{A}\right).\label{EFE4AA}
\end{eqnarray}

In the above, $\rho$,  $p_{r}$, $p_{t}$ and $q$ are the energy density, radial pressure, tangential pressure and radial heat flux respectively.

\subsection{The Karmarkar condition}

The Karmarkar condition \cite{kar}, which allows for the embedding of a four-dimensional spacetime into a five-dimensional pseudo-Euclidean space, is given in terms of the following relationship with respect to the components of the Riemann tensor, namely

\begin{equation} \label{kar1}
	\mathcal{R}_{1010} \mathcal{R}_{2323} = \mathcal{R}_{1212} \mathcal{R}_{3030} - \mathcal{R}_{1220} \mathcal{R}_{1330},
\end{equation}

where the notation $(0, 1, 2, 3)$ represents the coordinates $(t, r, \theta, \phi)$.
We then consider the metric (\ref{le.01}) and calculate Karmarkar's condition for a shearing, nonstatic spherically symmetric metric. The relevant nonzero Riemann tensor components are

\begin{eqnarray} \label{genriemann}
	\mathcal{R}_{1010} &=& \frac{1}{AB}\biggl(-B^2\ddot{B}A+A^2{A}^{\prime\prime}B-{B}^{\prime}{A}^{\prime}A^2+\dot{B}\dot{A}B^2\biggr),  \label{gr1010} \nonumber \\
	\mathcal{R}_{1212} &=& \frac{Y}{A^2 B}\biggl(-{Y}^{\prime\prime}BA^2+ {B}^{\prime}{Y}^{\prime}A^2+B^2\dot{B}\dot{Y}\biggr),	\label{gr1212} \nonumber \\
	\mathcal{R}_{1220} &=& \frac{-Y}{AB}\biggl(-{\dot{Y}}^{\prime}AB+\dot{B}{Y}^{\prime}A+ {A}^{\prime}\dot{Y}B\biggr), \label{gr1220} \nonumber \\
	\mathcal{R}_{1330} &=& \frac{-Y\sin^2{\theta}}{A B}\biggl(-{\dot{Y}}^{\prime}BA+\dot{B}{Y}^\prime A+{A}^{\prime}\dot{Y}B\biggr), \label{gr1330} \nonumber \\
	\mathcal{R}_{2323} &=& \frac{Y^2\sin^2{\theta}}{B^2 A^2}\biggl(B^2A^2-{{Y}^{\prime}}^2A^2+{\dot{Y}}^2B^2\biggr), \label{gr2323} \nonumber \\
	\mathcal{R}_{3030} &=& \frac{Y\sin^2{\theta}}{AB^2}\biggl(-\ddot{Y}B^2A+A^2{A}^{\prime}{Y}^{\prime}+\dot{Y}\dot{A}B^2\biggr). \label{gr3030} \nonumber \\
\end{eqnarray}

which results in the following expression for the Karmarkar condition:

\begin{eqnarray} \label{fullkarmarkarAA}
	0 &=& AB\bigg(A {Y}^{\prime}\dot{B} +B{A}^{\prime}\dot{Y}-AB\dot{Y}^{\prime}\bigg)^2 \nonumber \\
	&& + \bigg(A^2B^2-A^2{{Y}^{\prime}}^2+B^2{\dot{Y}}^2\bigg) \nonumber \\
	&& \times \bigg(-A^2{A}^{\prime}{B}^{\prime}+A^2B{A}^{\prime\prime}+B^2\dot{A}\dot{B}-AB^2\ddot{B}\bigg) \nonumber \\
	&& - \bigg(A^2{B}^{\prime}{Y}^{\prime}-A^2B{Y}^{\prime\prime}+B^2\dot{B}\dot{Y}\bigg)\nonumber \\
	&& \times \bigg(A^2{A}^{\prime}{Y}^{\prime}+B^2\dot{A}\dot{Y}-AB^2\ddot{Y}\bigg).
\end{eqnarray}

\section{Lie symmetries of differential equations}

In the context of geometry a differential equation (DE) may be considered as a function $H=H(y^{i},u^{A},u_{,i}^{A},u_{,ij}^{A},...)$ in the jet-space $B=B\left(  y^{i},u^{A},u_{,i}^{A},u_{,ij}^{A},...\right) $, where $\left\{y^{i}\right\}$ are the independent variables and $u^{A}$ are the dependent
variables, while comma means derivative with respect to the variable $y^{i}$, that is $u_{,i}^{A}=\frac{\partial u^{A}}{\partial y^{i}}$.

Consider the infinitesimal one-parameter point transformation

\begin{eqnarray}
\bar{x}^{i}  &=& x^{i}+\varepsilon\xi^{i}(x^{k},u^{B})~,\label{pr.01}\\
\bar{u}^{A}  &=& \bar{u}^{A}+\varepsilon\eta^{A}(x^{k},u^{B})~, \label{pr.02}%
\end{eqnarray}
with generator%
\begin{equation}
\mathbf{X}=\xi^{i}(x^{k},u^{B})\partial_{x^{i}}+\eta^{A}(x^{k},u^{B}%
)\partial_{u^{A}}~. \label{pr.03}%
\end{equation}

The vector field $\mathbf{X}$ which defines the infinitesimal transformation (\ref{pr.01})-(\ref{pr.02}) is called a Lie point symmetry of the DE $H$ if
there exists a function $\kappa$ such that the following condition holds \cite{Stephani,Bluman,ibra}

\begin{equation}
\mathbf{X}^{[n]}(H)=\kappa H~,~modH=0, \label{pr.04}%
\end{equation}
where
\begin{equation}
\mathbf{X}^{[n]}=\mathbf{X}+\eta_{i}^{A}\partial_{u_{i}^{A}}+\eta_{ij}%
^{A}\partial_{u_{ij}^{A}}+... \label{pr.05}%
\end{equation}
is the $n$th extension vector. Coefficient $\eta_{i}^{A}$ of the first extension vector is defined as

\begin{equation}
\eta_{i}^{A} = \eta_{,i}^{A}+u_{,i}^{B}\eta_{,B}^{A}-\xi_{,i}^{j}u_{,j}^{A}-u_{,i}^{A}u_{,j}^{B}\xi_{,B}^{j}~, \label{pr.06}%
\end{equation}
coefficient~$\eta_{ij}^{A}$ of the second extension vector is given by the expression

\begin{eqnarray}
\eta_{ij}^{A}  &=& \eta_{,ij}^{A}+2\eta_{,B(i}^{A}u_{,j)}^{B}-\xi_{,ij}^{k}u_{,k}^{A}+\eta_{,BC}^{A}u_{,i}^{B}u_{,j}^{C}-2\xi_{,(i|B|}^{k}u_{j)}^{B}u_{,k}^{A}\nonumber\\
&&  -\xi_{,BC}^{k}u_{,i}^{B}u_{,j}^{C}u_{,k}^{A}+\eta_{,B}^{A}u_{,ij}^{B}-2\xi_{,(j}^{k}u_{,i)k}^{A} \nonumber \\
&& -\xi_{,B}^{k}\left(  u_{,k}^{A}u_{,ij}^{B}+2u_{(,j}^{B}u_{,i)k}^{A}\right)  \label{pr.07}%
\end{eqnarray}

while coefficient $\eta_{ij...j_{n}}^{A}$ of the $n$th extension vector is defined as

\begin{equation}
\eta_{ij...j_{n}}^{A}=D\eta_{ij...j_{n-1}}^{A}-u_{ij..k}^{A}D_{j_{n}}\xi^{k}%
\end{equation}

The main application of Lie point symmetries of a DE is focused on the construction of invariant functions which can be used for the determination of
invariant solutions also known as similarity solutions.

For the Lie point symmetry $\mathbf{X}$ of the differential equation $H$ we define the Lagrange system \cite{Stephani,Bluman,ibra} as

\begin{equation}
\frac{dx^{i}}{\xi^{i}}=\frac{du^{A}}{\eta^{A}}=\frac{du_{i}^{A}}{\eta_{\left[i\right]  }^{A}}=\frac{du_{ij}^{A}}{\eta_{\left[  ij\right]  }^{A}}=...
\end{equation}

whose solution provides the characteristic functions \\
$W^{\left[  0\right]}\left(  y^{k},u\right)  ,~W^{\left[  1\right]  }\left(  y^{k},u,u_{i}\right)
$, $W^{\left[  2\right]  }\left(  y^{k},u,u_{,i},u_{ij}\right)  ,...~$. \\
The characteristic functions, can be applied to reduce the order of the DE (in the case of ordinary differential equations) or the number of the dependent
variables (in the case of partial differential equations).

\subsection{One-dimension optimal system}

For a given differential equation $H$ which admit a Lie algebra $G_{n}$ of dimension $\dim G_{n}=n$ and elements $\left\{  X_{1},~X_{2},~...~X_{n}\right\}  ,$ we consider the two generic vector fields \cite{olver}

\begin{eqnarray}\label{sw.04}
Z &= \sum\limits_{i=1}^{n}a_{i}X_{i}~, \hspace{0.5cm} ~W =& \sum\limits_{i=1}^{n}b_{i}X_{i}~,
\end{eqnarray}

where $a_{i},~b_{i}$ are constants.

The vector fields $Z,~W$ are equivalent and leads to the same similarity
transformation if and only if

\begin{equation}
\mathbf{W} = {\displaystyle\prod\limits_{i}}Ad\left(  \exp\left(  \varepsilon_{i}X_{i}\right)  \right)  \mathbf{Z} \label{sw.05}
\end{equation}

or

\begin{equation}
W=cZ~,~c=const. \label{sw.06}
\end{equation}

where the operator $Ad\left(  \exp\left(  \varepsilon X_{i}\right)  \right) $ is the Adjoint operator defined as \cite{olver}

\begin{equation}
Ad\left(  \exp\left(  \varepsilon X_{i}\right)  \right)  X_{j} = X_{j} - \varepsilon\left[  X_{i},X_{j}\right]  +\frac{1}{2}\varepsilon^{2}\left[
X_{i},\left[  X_{i},X_{j}\right]  \right]  +... \label{sw.07}
\end{equation}

which is used to determine the Adjoint representation. Hence, in order to perform a complete classification for the similarity solutions of a given differential equation we should determine all the one-dimensional independent symmetry vectors of the Lie algebra $G_{n}$.

In the following sections we consider special forms for the unknown metric functions such that there is only one unknown function, for that models we perform a detailed analysis of the Karmarkar condition by using the Lie's theory. In particular we determine the Lie point symmetries and the one-dimensional optimal system for the Karmarkar condition, while we determine similarity solutions.

\section{Model A: Conformally flat metric}

Consider $A\left(  t,r\right)  = B\left(  t,r\right)  $ and $Y\left(t,r\right)  =rB\left(  t,r\right) $. In this case, the line element (\ref{le.01}) is,

\begin{equation}
ds^2 = B^2 \left(  t,r\right)  \left( -dt^{2}+dr^{2}+r^{2}d\Omega^{2}\right) . \label{sw.08}
\end{equation}

A spacetime with line element (\ref{sw.08}) is conformally flat and admits fifteen Conformal Killing vector fields (CKVs). For the conformally flat metric (\ref{sw.08}) where $B\left(  t,r\right) $ is the unique unknown function, the Karmarkar condition becomes,

\begin{eqnarray} \label{sw.09}
0 &=& -4 r^2 \dot{B}B' \dot{B}' + B \left(r^2 \dot{B}'^2 + B'^2 - r B'' B' + r \ddot{B}\left(B' - r B'' \right)\right) \nonumber \\
&& +2 r B'^2 \left(r \ddot{B} + B' \right) + 2 r \dot{B}^2 \left(r B'' - B' \right)
\end{eqnarray}

We apply the Lie theory to equation (\ref{sw.09}) from which we obtain the Lie point symmetry vectors

\[
X_{1}=\partial_{t}~,~X_{2}=B\partial_{B}~,~X_{3}=B^{2}\partial_{B}~,~X_{4}=\frac{1}{r}\partial_{r}~,~
\]
\[
X_{5}=tB^{2}\partial_{B}~,~X_{6}=\left(  r^{2}-t^{2}\right)  B^{2}\partial_{B}~,~X_{7}=t\partial_{t}+r\partial_{r}.
\]

The admitted Lie symmetries form a seven-dimensional Lie algebra $G_{A}$, i.e. $\dim G_{A} = 7$, and the associated commutators are shown in Table \ref{tab1}.
Moreover, in Table \ref{tab2} we present the Adjoint representation for the elements of the Lie algebra $G_{A}$.

We continue by using the results in Tables \ref{tab1} and \ref{tab2} to derive the one-dimensional optimal system for the partial differential
equation (\ref{sw.09}). Consider the generic symmetry vector

\begin{equation}
Z=\alpha_{1}X_{1}+\alpha_{2}X_{2}+\alpha_{3}X_{3}+\alpha_{4}X_{4}+\alpha_{5}X_{5}+\alpha_{6}X_{6}+\alpha_{7}X_{7}%
\end{equation}

From Table \ref{tab2} we see that by applying the following adjoint representations

\begin{eqnarray}
Z^{\prime} &=& Ad\left(  \exp\left(  \varepsilon_{2}X_{2}\right)  \right)Ad\left(  \exp\left(  \varepsilon_{5}X_{5}\right)  \right)  Ad\left(
\exp\left(  \varepsilon_{6}X_{6}\right)  \right)  \times \nonumber \\
&& Ad\left(  \exp\left(\varepsilon_{1}X_{1}\right)  \right)  Z,
\end{eqnarray}

where for specific values of $\varepsilon_{1,}~\varepsilon_{2},~\varepsilon_{5}$ and $\varepsilon_{6}$ it follows

\begin{equation}
Z^{\prime}=\alpha_{2}^{\prime}X_{2}+\alpha_{4}^{\prime}X_{4}+\alpha_{7}^{\prime}X_{7}.
\end{equation}

Thus, the two vector fields $Z^{\prime}$ and $Z$ are equivalent and lead to the same similarity solution. Coefficient constants $\alpha_{2},~\alpha_{4}$
and $\alpha_{7}$ are called relative invariants of the full adjoint action.
Thus, in order to derive the relative invariants we solve the following system of partial differential equation \cite{olver}

\begin{equation}
\Delta\left(  \phi\left(  \alpha_{i}\right)  \right)  = C_{ij}^{k}\alpha^{i}\frac{\partial}{\partial\alpha_{j}} \label{ss.01}
\end{equation}

where $C_{ij}^{k}$ are the structure constants of the Lie algebra $G_{A}$ as presented in \ref{tab1}. Hence, system (\ref{ss.01}) becomes

\begin{eqnarray}
\alpha_{5}\frac{\partial\phi}{\partial\alpha_{3}}-2\alpha_{6}\frac
{\partial\phi}{\partial\alpha_{5}} + \alpha_{7}\frac{\partial\phi}{\partial\alpha_{1}} &  = 0,\\
\alpha_{3}\frac{\partial\phi}{\partial\alpha_{3}}+\alpha_{5}\frac{\partial
\phi}{\partial\alpha_{5}}+\alpha_{6}\frac{\partial\phi}{\partial\alpha_{6}} &
=0,\\
-\alpha_{2}\frac{\partial\phi}{\partial\alpha_{3}} &=& 0,\\
2\alpha_{6}\frac{\partial\phi}{\partial\alpha_{3}} &=& 0,\\
-\alpha_{1}\frac{\partial\phi}{\partial\alpha_{3}}-\left(  \alpha_{2}%
+\alpha_{7}\right)  \frac{\partial\phi}{\partial\alpha_{5}} &=& 0,\\
2\alpha_{1}\frac{\partial\phi}{\partial\alpha_{5}}-\left(  \alpha_{2}%
+2\alpha_{7}\right)  \frac{\partial\phi}{\partial\alpha_{6}}-2\alpha_{4}%
\frac{\partial\phi}{\partial\alpha_{3}} &=& 0,\\
-\alpha_{1}\frac{\partial\phi}{\partial\alpha_{1}}+\alpha_{5}\frac
{\partial\phi}{\partial\alpha_{5}}+\alpha_{6}\frac{\partial\phi}%
{\partial\alpha_{6}} &=& 0,
\end{eqnarray}

from where it follows $\phi\left(  \alpha_{i}\right)  = \phi\left(  \alpha_{2},\alpha_{4},\alpha_{7}\right)  $, that is, the relative invariants are
$\left\{  \alpha_{2},\alpha_{4},\alpha_{7}\right\} $.

For $\alpha_{7}=0$, we find the equivalent symmetry vector

\begin{equation}
Z^{\prime\prime}=\alpha_{1}^{\prime\prime}X_{1}+\alpha_{2}^{\prime\prime}%
X_{2}+\alpha_{4}^{\prime\prime}X_{4}.
\end{equation}
On the other hand for $\alpha_{2}=0$, the equivalent symmetry vector is
\begin{equation}
Z^{\prime\prime\prime}=\alpha_{4}^{\prime\prime\prime}X_{4}+\alpha_{7}%
^{\prime\prime\prime}X_{7}.
\end{equation}
while when $\alpha_{4}=0$, the resulting equivalent symmetry vector is derived%
\begin{equation}
Z^{\prime\prime\prime\prime}=\alpha_{2}^{\prime\prime\prime\prime}X_{2}%
+\alpha_{7}^{\prime\prime\prime\prime}X_{7}.
\end{equation}

Similarly, the following equivalent symmetry vectors are obtained: 

\[
\{\alpha_{2} = 0,~\alpha_{4}=0\} ~\Rightarrow~ \bar{Z}=\bar{\alpha}_{7}X_{7}~, 
\]
\[
\{\alpha_{4}=0,~\alpha_{7}=0\} ~\Rightarrow~ \bar{Z}^{\prime}=\bar{\alpha}_{1}^{\prime}X_{1}+\bar{\alpha}_{2}^{\prime}X_{2}~, 
\]
\[
\{\alpha_{2}=0,~\alpha_{7}=0\} ~\Rightarrow~ \bar{Z}^{\prime\prime}=\bar{\alpha}_{1}^{\prime\prime}X_{1}+\bar{\alpha}_{4}^{\prime\prime}X_{4}+\bar{\alpha}_{5}^{\prime\prime}X_{5}+\bar{\alpha}_{6}^{\prime\prime}X_{6}~, 
\]
\[
\{\alpha_{2}=0,~\alpha_{4}=0,~\alpha_{7}=0\} ~\Rightarrow~ \bar{Z}^{\prime\prime\prime} = \bar{\alpha}_{1}^{\prime\prime\prime}X_{1}+\bar{\alpha}_{5}^{\prime\prime\prime}X_{5}+\bar{\alpha}_{6}^{\prime\prime\prime}X_{6}. 
\]

Thus, a one-dimensional optimal system is constructed from the one-dimensional Lie algebras:

\[
\{X_1\}  ,~ \{X_2\}  ,~ \{X_3\} ,~ \{X_4\} ,~ \{X_5\} ,~ \{X_6\} ,~ \{X_7\} ,~
\]
\[
\{X_1 + \alpha X_5\}  ,~ \{X_1 + \alpha X_6\}  ,~ \{X_5 + \alpha X_6\} ,~ \{X_2 + \alpha X_7\} ,~
\]
\[
\{X_5 + \alpha X_7\}  ,~ \{X_1 + \alpha X_4\}  ,~ \{X_4 + \alpha X_5\} ,~ \{X_4 + \alpha X_6\} ,~
\]
\[
\{X_1 + \alpha X_2 + \beta X_4\}  ,~ \{X_1 + \alpha X_5 + \beta X_6\}  ,~ \{X_4 + \alpha X_1 + \beta X_5\} ,~ 
\]
\[
\{X_4 + \alpha X_1 + \beta X_6\}  ,~ \{X_4 + \alpha X_5 + \beta X_6\}  ,~ \{X_2 + \alpha X_4 + \beta X_7\} .
\]

We proceed by using the Lie symmetries to determine similarity transformations, in the following we present the application of Lie point symmetries which lead to exact solutions expressed in closed-form functions.

\begin{table}[tbp] \centering
\caption{Commutators of the admitted Lie point symmetries for the Karmarkar condition (\ref{sw.08})}
\begin{tabular}
[c]{cccccccc}\hline\hline
$\left[  ~,~\right]  $ & $\mathbf{X}_{1}$ & $\mathbf{X}_{2}$ & $\mathbf{X}_{3}$ & $\mathbf{X}_{4}$ & $\mathbf{X}_{5}$ & $\mathbf{X}_{6}$ &
$\mathbf{X}_{7}$\\
$\mathbf{X}_{1}$ & $0$ & $0$ & $0$ & $0$ & $X_{3}$ & $-2X_{5}$ & $X_{1}$\\
$\mathbf{X}_{2}$ & $0$ & $0$ & $X_{3}$ & $0$ & $X_{5}$ & $X_{6}$ & $0$\\
$\mathbf{X}_{3}$ & $0$ & $-X_{3}$ & $0$ & $0$ & $0$ & $0$ & $0$\\
$\mathbf{X}_{4}$ & $0$ & $0$ & $0$ & $0$ & $0$ & $2X_{3}$ & $0$\\
$\mathbf{X}_{5}$ & $-X_{3}$ & $-X_{5}$ & $0$ & $0$ & $0$ & $0$ & $-X_{5}$\\
$\mathbf{X}_{6}$ & $2X_{5}$ & $-X_{6}$ & $0$ & $-2X_{3}$ & $0$ & $0$ &
$-2X_{6}$\\
$\mathbf{X}_{7}$ & $-X_{1}$ & $0$ & $0$ & $0$ & $X_{5}$ & $2X_{6}$ &
$0$\\\hline\hline
\end{tabular}
\label{tab1}
\end{table}

\begin{table*}[tbp] \centering
\caption{Adjoint representation for the admitted Lie point symmetries for the Karmarkar condition (\ref{sw.08})}
\begin{tabular}
[c]{cccccccc}\hline\hline
$Ad\left(  e^{\left(  \varepsilon\mathbf{X}_{i}\right)  }\right)
\mathbf{X}_{j}$ & $\mathbf{X}_{1}$ & $\mathbf{X}_{2}$ & $\mathbf{X}_{3}$ &
$\mathbf{X}_{4}$ & $\mathbf{X}_{5}$ & $\mathbf{X}_{6}$ & $\mathbf{X}_{7}%
$\\\hline
$\mathbf{X}_{1}$ & $X_{1}$ & $X_{2}$ & $X_{3}$ & $X_{4}$ & $X_{5}-\varepsilon
X_{3}$ & $X_{5}+2\varepsilon X_{5}-\varepsilon^{2}X_{3}$ & $X_{7}-\varepsilon
X_{1}$\\
$\mathbf{X}_{2}$ & $X_{1}$ & $X_{2}$ & $e^{-\varepsilon}X_{3}$ & $X_{4}$ &
$e^{-\varepsilon}X_{5}$ & $e^{-\varepsilon}X_{6}$ & $X_{7}$\\
$\mathbf{X}_{3}$ & $X_{1}$ & \thinspace$X_{2}+\varepsilon X_{3}$ & $X_{3}$ &
$X_{4}$ & $X_{5}$ & $X_{6}$ & $X_{7}$\\
$\mathbf{X}_{4}$ & $X_{1}$ & $X_{2}$ & $X_{3}$ & $X_{4}$ & $X_{5}$ &
$X_{6}-2\varepsilon X_{3}$ & $X_{7}$\\
$\mathbf{X}_{5}$ & $X_{1}+\varepsilon X_{3}$ & $X_{2}+\varepsilon X_{5}$ &
$X_{3}$ & $X_{4}$ & $X_{5}$ & $X_{6}$ & $X_{7}+\varepsilon X_{5}$\\
$\mathbf{X}_{6}$ & $X_{1}-2\varepsilon X_{5}$ & $X_{2}+\varepsilon X_{6}$ &
$X_{3}$ & $X_{4}+2\varepsilon X_{3}$ & $X_{5}$ & $X_{6}$ & $X_{7}+2\varepsilon
X_{6}$\\
$\mathbf{X}_{7}$ & $e^{\varepsilon}X_{1}$ & $X_{2}$ & $X_{3}$ & $X_{4}$ &
$e^{-\varepsilon}X_{5}$ & $e^{-2\varepsilon}X_{6}$ & $X_{7}$\\\hline\hline
\end{tabular}
\label{tab2}
\end{table*}

\section{Model B: Weyl-free collapse}

We proceed with our study on the Karmarkar condition by assuming Weyl-free collapse spacetimes, where

\begin{eqnarray}
A\left(  t,r\right) &=&\left(  1-\omega^{2}r^{2}\right)^{1/2} B\left(  t,r\right) \nonumber \\
Y\left(  t,r\right) &=& rB\left(  t,r\right)
\end{eqnarray}

that is, the line element (\ref{le.01}) becomes

\begin{equation}
ds^{2}=B^{2}\left(  t,r\right)  \left(  -\left(  1-\omega^{2}r^{2}\right) dt^{2}+dr^{2}+r^{2}d\Omega^{2}\right)
\end{equation}

with $\omega\neq0$. Recall that in the limit where $\omega=0,$ the latter line element describes the conformally flat spacetime (\ref{sw.08}). The Karmarkar condition for the above line element is

\begin{eqnarray}
0 &=& r \omega ^2 B^2 \left(\left(r^2 \omega ^2+1\right) B' + r \left(r^2 \omega ^2-1\right) B'' \right) \nonumber \\
&& + B \bigg(r^2 \dot{B}' \left(\left(1-r^2 \omega ^2\right) \dot{B}' + 2 r \omega ^2 \dot{B}\right) + r \left(r^2 \omega ^2-1\right)B' B'' \nonumber \\
&& + r \left(r^2 \omega^2 - 1\right)\left(r B'' - B' \right)\ddot{B} + \left(1 + r^2 \omega ^2 - r^4 \omega ^4\right) B'^2\bigg) \nonumber \\
&& - 2 r \bigg(\dot{B}^2 \left(\left(r^2 \omega ^2+1\right) B' + r \left(r^2 \omega ^2-1\right) B'' \right)  \nonumber \\
&& - 2 r \left(r^2 \omega ^2-1\right) B' \dot{B}' \dot{B} + \left(r^2 \omega ^2-1\right)\left(r \ddot{B} + B' \right)B'^2 \bigg).
\end{eqnarray}

This admits a six-dimensional Lie algebra, $G_{B}$, consisted by the Lie symmetry vectors

\[
Y_{1}=\partial_{t}~,~Y_{2}=B\partial_{B}~,~Y_{3}=\left(  1+\omega^{2}r^{2}\right)  B^{2}\partial_{B}~,
\]%
\[
Y_{4}=e^{-2\omega t}\left(  \omega^{2}r^{2}-1\right)  \partial_{B}~,~Y_{5}=e^{2\omega t}\left(  \omega^{2}r^{2}-1\right)  \partial_{B}~,~
\]%
\[
Y_{6}=\left(  \omega^{2}r^{2}-1\right)  \left(  2\omega^{2}B\partial_{B}-\frac{\left(  1+\omega^{2}r^{2}\right)  }{r}\partial_{r}\right)  .
\]

In Table \ref{tab3} we present the commutators of the Lie algebra $G_{B}$ while in Table \ref{tab4} the Adjoint representation is given, necessary for
the derivation of the one-dimensional optimal system.

Hence, the system of the partial differential equations (\ref{ss.01}) which provides the relative invariants is simplified as

\begin{equation}
\frac{\partial\phi}{\partial\alpha_{3}} = 0~~,~~\frac{\partial\phi}{\partial\alpha_{4}}=0~~,~~\frac{\partial\phi}{\partial\alpha_{5}}=0~,~
\end{equation}

which means that $\phi\left(  \alpha_{I}\right)  =\phi\left( \alpha_{1},\alpha_{2},\alpha_{6}\right) $.

Assume the generic symmetry vector $W=\alpha_{1}Y_{1}+\alpha_{2}Y_{2}+\alpha_{3}Y_{3}+\alpha_{4}Y_{4}+\alpha_{5}Y_{5}+\alpha_{6}Y_{6}$, then when
$\alpha_{1}\alpha_{2}\alpha_{6}\neq0$, the equivalent symmetry vector is

\begin{equation}
W^{\prime}=\alpha_{1}^{\prime}Y_{1}+\alpha_{2}^{\prime}Y_{2}+\alpha_{6}^{\prime}Y_{6}.
\end{equation}

For $\alpha_{6}=0$, the equivalent vector field is $W^{\prime\prime}=\alpha_{1}^{\prime\prime}Y_{1}+\alpha_{2}^{\prime\prime}Y_{2},~$for
$\alpha_{4}=0$, it follows $W^{\prime\prime\prime}=\alpha_{1}^{\prime\prime\prime}Y_{1}+\alpha_{3}^{\prime\prime\prime}Y_{3}+\alpha_{6}^{\prime\prime\prime}Y_{6},~$while when $\alpha_{1}=0$ we have the equivalent symmetry vector $~W^{\prime\prime\prime\prime}=\alpha_{2}^{\prime\prime\prime\prime}Y_{2}+\alpha_{6}^{\prime\prime\prime\prime}Y_{6}.~$

Similarly, the following equivalent symmetry vectors are obtained: 

\[
\{\alpha_{1} = 0,~\alpha_{6}=0\} ~\Rightarrow~ \bar{W}=\bar{\alpha}_{2}Y_{2}~, 
\]
\[
\{\alpha_{1}=0,~\alpha_{2}=0\} ~\Rightarrow~ \bar{W}^{\prime}=\bar{\alpha}_{3}^{\prime}Y_{3}+\bar{\alpha}_{6}^{\prime}Y_{6}~, 
\]
\[
\{\alpha_{2}=0,~\alpha_{6}=0\} ~\Rightarrow~ \bar{W}^{\prime\prime}=\bar{\alpha}_{1}^{\prime\prime}Y_{1}+\bar{\alpha}_{3}^{\prime\prime}Y_{3}~, 
\]
\[
\{\alpha_{1}=0,~\alpha_{2}=0,~\alpha_{6}=0\} ~\Rightarrow~ \bar{W}^{\prime\prime\prime}=\bar{\alpha}_{3}^{\prime\prime\prime}Y_{3}+\bar{\alpha}_{4}^{\prime\prime\prime}Y_{4}+\bar{\alpha}_{5}^{\prime\prime\prime}Y_{5}. 
\]

We conclude that the one-dimensional optimal system consisted by the one-dimensional Lie algebras:

\[
\{Y_1\} ~,~ \{Y_2\} ~,~ \{Y_3\} ~,~ \{Y_4\} ~,~ \{Y_5\} ~,~ \{Y_6\} ~,
\]  
\[
\{Y_1 + \alpha Y_2\}  ,~ \{Y_1 + \alpha Y_3\}  ,~ \{Y_1 + \alpha Y_6\} ,~ \{Y_2 + \alpha Y_6\} ,~
\]
\[
\{Y_3 + \alpha Y_4\}  ,~ \{Y_3 + \alpha Y_5\}  ,~ \{Y_3 + \alpha Y_6\} ,~ \{Y_4 + \alpha Y_5\} ,~
\]
\[
 \{Y_1 + \alpha Y_3 + \alpha Y_6\}  ,~ \{Y_1 + \alpha Y_2 + \beta Y_6\} ,~ \{Y_3 + \alpha Y_4 + \beta Y_5\}.
\]

We proceed with the presentation of similarity solutions which are expressed by closed-form functions.

\begin{table}[tbp] \centering
\caption{Commutators of the admitted Lie point symmetries for the Karmarkar condition for the Weyl-free collapse spacetimes}
\begin{tabular}
[c]{ccccccc}\hline\hline
$\left[  ~,~\right]  $ & $\mathbf{Y}_{1}$ & $\mathbf{Y}_{2}$ & $\mathbf{Y}_{3}$ & $\mathbf{Y}_{4}$ & $\mathbf{Y}_{5}$ & $\mathbf{Y}_{6}$\\
$\mathbf{Y}_{1}$ & $0$ & $0$ & $0$ & $-2\omega Y_{4}$ & $2\omega Y_{5}$ &
$0$\\
$\mathbf{Y}_{2}$ & $0$ & $0$ & $Y_{3}$ & $Y_{4}$ & $Y_{5}$ & $0$\\
$\mathbf{Y}_{3}$ & $0$ & $-Y_{3}$ & $0$ & $0$ & $0$ & $0$\\
$\mathbf{Y}_{4}$ & $2\omega Y_{4}$ & $-Y_{4}$ & $0$ & $0$ & $0$ & $4\omega
^{2}Y_{4}$\\
$\mathbf{Y}_{5}$ & $-2\omega Y_{5}$ & $-Y_{5}$ & $0$ & $0$ & $0$ &
$4\omega^{2}Y_{5}$\\
$\mathbf{Y}_{6}$ & $0$ & $0$ & $0$ & $-4\omega^{2}Y_{4}$ & $-4\omega^{2}Y_{5}$
& $0$\\\hline\hline
\end{tabular}
\label{tab3}
\end{table}

\begin{table*}[tbp] \centering
\caption{Adjoint representation for the admitted Lie point symmetries for the Karmarkar condition for the Weyl-free collapse spacetimes}
\begin{tabular}
[c]{ccccccc}\hline\hline
$Ad\left(  e^{\left(  \varepsilon\mathbf{Y}_{i}\right)  }\right)
\mathbf{Y}_{j}$ & $\mathbf{Y}_{1}$ & $\mathbf{Y}_{2}$ & $\mathbf{Y}_{3}$ &
$\mathbf{Y}_{4}$ & $\mathbf{Y}_{5}$ & $\mathbf{Y}_{6}$\\\hline
$\mathbf{Y}_{1}$ & $Y_{1}$ & $Y_{2}$ & $Y_{3}$ & $e^{2\omega\varepsilon}Y_{4}$
& $e^{-2\omega\varepsilon}Y_{5}$ & $Y_{6}$\\
$\mathbf{Y}_{2}$ & $Y_{1}$ & $Y_{2}$ & $e^{-\varepsilon}Y_{3}$ &
$e^{-\varepsilon}Y_{4}$ & $e^{-\varepsilon}Y_{5}$ & $Y_{6}$\\
$\mathbf{Y}_{3}$ & $Y_{1}$ & $Y_{2}+\varepsilon Y_{3}$ & $Y_{3}$ & $Y_{4}$ &
$Y_{5}$ & $Y_{6}$\\
$\mathbf{Y}_{4}$ & $Y_{1}-2\omega\varepsilon Y_{4}$ & $Y_{2}+\varepsilon
Y_{4}$ & $Y_{3}$ & $Y_{4}$ & $Y_{5}$ & $Y_{6}-4\omega^{2}\varepsilon Y_{4}$\\
$\mathbf{Y}_{5}$ & $Y_{1}+2\omega\varepsilon Y_{5}$ & $Y_{2}+\varepsilon
Y_{5}$ & $Y_{3}$ & $Y_{4}$ & $Y_{5}$ & $Y_{6}-4\omega^{2}\varepsilon Y_{5}$\\
$\mathbf{Y}_{6}$ & $Y_{1}$ & $Y_{2}$ & $Y_{3}$ & $e^{4\omega^{2}\varepsilon
}Y_{4}$ & $e^{4\omega^{2}\varepsilon}Y_{5}$ & $Y_{6}$\\\hline\hline
\end{tabular}
\label{tab4}
\end{table*}

\section{Model C: Shear-free collapse}

Consider the shear-free collapse spacetime where $Y\left(  t,r\right) = rB\left(  t,r\right)  $ and $A\left(  t,r\right)  = B\left(  t,r\right)^{-N}$ with $N\neq-1$. Hence, the line element (\ref{le.01}) reads

\begin{equation}
ds^{2}=B^{2}\left(  t,r\right)  \left(  -B^{-2-2N}\left(  t,r\right)
dt^{2}+dr^{2}+r^{2}d\Omega^{2}\right)  .
\end{equation}

The Karmarkar condition then gives

\begin{eqnarray}
0 &=& \left(2 n^2+3 n+1\right) r^2 \dot{B}^2 B'^2 B^{2n+2} - n (n+1) r^2 B'^4  \nonumber \\
&& - 2 r \left(\dot{B}^2 \left(n r B'' - n B' \right) - (n-1) r B' \dot{B}' \dot{B} - r \ddot{B} B'^2\right)B^{2 n+3} \nonumber \\
&& + r \left(r \dot{B}'^2 + \ddot{B}\left(B' - r B'' \right)\right) B^{2 n+4} + n B' \left(r B'' - B' \right) B^2 \nonumber \\
&& - 2 n(n+2) r B'^3 B.
\end{eqnarray}

The resulting Karmarkar condition admits a three dimensional Lie algebra, $G_{C}$, consisted by the Lie symmetry vectors

\begin{equation}
Z_{1} = \partial_{t}~,~Z_{2}=r\partial_{r}-\frac{1}{N+1}B\partial_{B}~,~Z_{3}=t\partial_{t}+\frac{1}{N+1}B\partial_{B},~
\end{equation}

with commutators and Adjoint representation as given in Table \ref{tab5}.

\begin{table*}[tbp] \centering
\caption{Commutators and Adjoint representation of the admitted Lie point symmetries for the Karmarkar condition for the Model C}
\begin{tabular}
[c]{ccccccccccccccc}\hline\hline
$\left[  ~,~\right]  $ & $\mathbf{Z}_{1}$ & $\mathbf{Z}_{2}$ & $\mathbf{Z}_{3}$ &  &  &  &  &  &  &  & $Ad\left(  e^{\left(  \varepsilon\mathbf{Z}%
_{i}\right)  }\right)  \mathbf{Z}_{j}$ & $\mathbf{Z}_{1}$ & $\mathbf{Z}_{2}$ &
$\mathbf{Z}_{3}$\\
$\mathbf{Z}_{1}$ & $0$ & $0$ & $Z_{1}$ &  &  &  &  &  &  &  & $\mathbf{Z}_{1}$
& $Z_{1}$ & $Z_{2}$ & $Z_{3}-\varepsilon Z_{1}$\\
$\mathbf{Z}_{2}$ & $0$ & $0$ & $0$ &  &  &  &  &  &  &  & $\mathbf{Z}_{2}$ &
$Z_{1}$ & $Z_{2}$ & $Z_{3}$\\
$\mathbf{Z}_{3}$ & $-Z_{1}$ & $0$ & $0$ &  &  &  &  &  &  &  & $\mathbf{Z}%
_{3}$ & $e^{\varepsilon}Z_{1}$ & $Z_{2}$ & $Z_{3}$\\\hline\hline
\end{tabular}
\label{tab5}
\end{table*}

Easily we calculate that the relative invariants are $\alpha_{2},\alpha_{3}$, from where we conclude that the one-dimensional optimal system consisted by
the one-dimensional Lie algebras

\[
\left\{  Z_{1}\right\}  ~,~\left\{  Z_{2}\right\}  ~\,,~\left\{
Z_{3}\right\}  ~,~\left\{  Z_{1}+\alpha Z_{2}\right\}  ~,~\left\{
Z_{2}+\alpha Z_{3}\right\}
\]

\section{A radiating, Weyl-free model}

By assuming Weyl-free collapse spacetimes where the line element is given by

\begin{equation}
ds^{2}=B^{2}\left(  t,r\right)  \left[  -\left(  1-\omega^{2}r^{2}\right) dt^{2}+dr^{2}+r^{2}d\Omega^{2}\right],
\end{equation}

application of the Lie symmetry vector $X_{1}+\alpha X_{6}$ gives the similarity solution

\begin{equation}
B\left( t, r\right)  = -\frac{8\alpha\omega^{2}}{\omega^{2}r^{2}+1}\left(\lambda_{1}e^{-4z\alpha\omega^{2}}-\lambda_{0}\right)  ^{-1}
\end{equation}

where $z=-\frac{1}{4\alpha\omega^{2}}\left(  \ln\left(  \frac{\omega^{2}r^{2}-1}{\omega^{2}r^{2}+1}\right)  +4\alpha\omega^{2}t\right) $.

By setting $\lambda_0 = 0$, we find that the solution satisfies the heat-flux boundary condition,

\begin{equation}
p_r = qB|_\Sigma
\end{equation}

and a simple radiating model may be generated. From the boundary condition constraint, an expression for $\alpha$ is obtained,

\begin{equation}
\alpha = \frac{\sqrt{3}\sqrt{2 + R^2\omega^2} - R\omega}{4\omega \sqrt{1 - R^2\omega^2}}
\end{equation}

where $R = r_\Sigma$ is the co-moving boundary. We examine the mass function obtained by calculating,

\begin{equation} \label{mass}
m = \frac{r^3 B \dot{B}^2}{2 A^2} - r^2 B' - \frac{r^3 B'^2}{2B}
\end{equation}

and then search for a function $\omega(t)$ such that the mass is constant during the distant past and then decreases monotonically from some point in time as the stellar object undergoes non-adiabatic gravitational collapse.

During this investigation, a definitive form for the function $\omega(t)$ was not obtainable and an ad hoc approach was used in establishing an approximate relationship. This involved calculating values for $\omega(t)$ for $t < t_i$ such that the mass function remained constant. The function

\begin{equation} \label{omega}
\omega (t) = \frac{a}{b t - c} + \frac{d}{t - e}
\end{equation}

was found to be appropriate for approximating the data obtained. \\
We consider a mass of $3 M_\odot$ prior to collapse $(t < -1000)$ with a comoving boundary $r_\Sigma = 3 \times 10^{10}cm / c = 1.00 s$. The following parameters were then obtained for the function (\ref{omega}) by solving (\ref{mass}) for time $t$, obtaining data for $\omega (t)$ at early times $(t < -1000)$ and then fitting the data to the function (\ref{omega}):

\begin{eqnarray}
 (a, b, c) &\rightarrow& (-0.727479, 0.191566, 1399.22) \nonumber \\
 (d, e) &\rightarrow& (-6.25399, 136.291)
\end{eqnarray}

Now that $\omega (t)$ has been determined, the gravitational potentials are fully specified and a radiating model, based on the Karmarkar condition with vanishing Weyl stress, has been generated.

\section{Discussion}

The solutions obtained for the conformally flat metric (Model A) did not offer the possibility for closed systems satisfying the heat flux boundary condition. Pressure isotropy is however easily invoked which can assist in setting parameters. Cosmological models might then be possible, to be explored in future work. \\
A solution for the Weyl-free collapse metric (Model B) was found to satisfy the heat-flux boundary condition and this was exploited in the previous section. The resulting mass function is shown in Figure 1. The radial profile is typical of uniform density matter and most of the mass is radiated as the collapse proceeds. The energy density, pressure and heat flux are shown in Figures 2 - 4. The energy density is uniform and relatively low for a stellar collapse process. The pressure behaves in a similar manner. We note that the heat flux increases towards the surface boundary as the collapse proceeds, so that little heat is generated near the gravitating centre.

\begin{figure}
\centering
\includegraphics[scale=0.85]{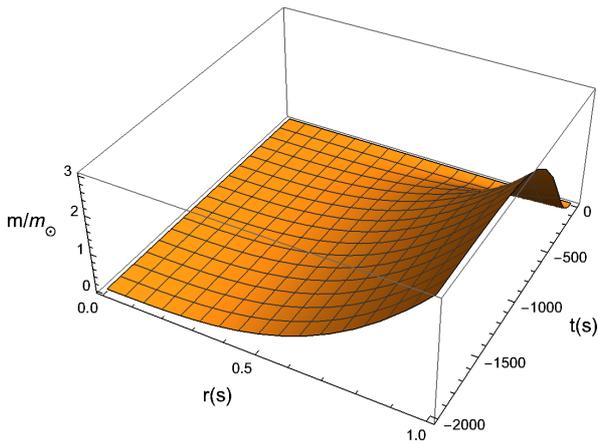}\caption{Mass radial-temporal profile}
\end{figure}

\begin{figure}
\centering
\includegraphics[scale=0.85]{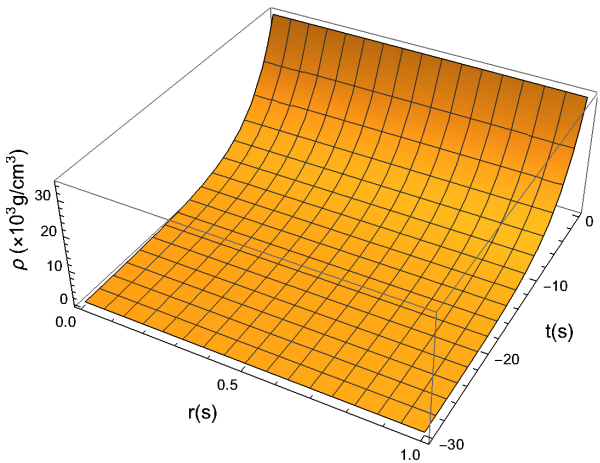}\caption{Energy density}
\end{figure}

\begin{figure}
\centering
\includegraphics[scale=0.85]{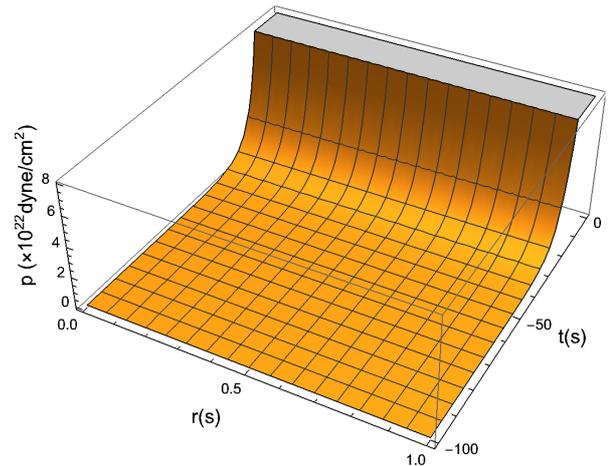}\caption{Pressure}
\end{figure}

\begin{figure}
\centering
\includegraphics[scale=0.85]{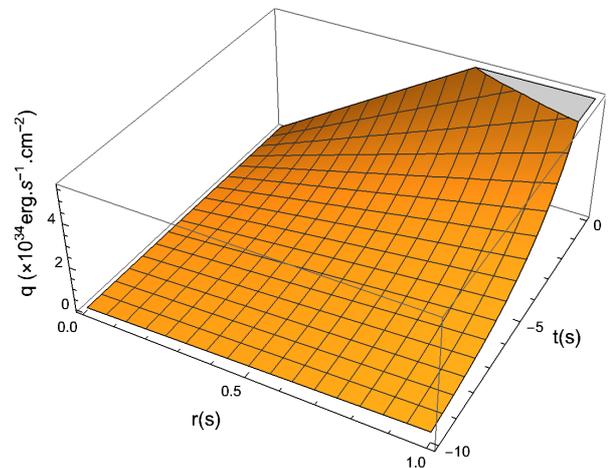}\caption{Heat flux}
\end{figure}

The shear-free solution (Model C) is yet to be explored and forms a basis for future work.

\section{Conclusion}

The use of Lie symmetries in obtaining solutions for conformally flat, Weyl-free and shear-free metrics which in addition, satisfy the Karmarkar condition has been shown. The solutions range from those that are almost trivial to those that are fairly complex. In searching for solutions that are consistent with the heat-flux boundary condition given according to Santos, a solution obtained for the Weyl-free collapse scenario was suitable. Other solutions from the other models did not offer a clear means in maintaining a time-independent heat-flux boundary condition. Some of these solutions may well be suited to cosmological models in which the energy density is related to the fourth power of the temperature.
In particular, a solution was found with respect to Weyl-free collapse which allows for a radiating model to be constructed. We see from Figures 1 - 4 that the physical parameters behave in a physically viable manner. The mass decreases monotonically, in a way that is similar to other gravitational collapse models. We note that the mass does not display a simple linear time-dependence as shown in other studies \cite{nai2}.
The solutions presented here compliment the recent solution obtained for the temporal evolution of a conformally flat interior, matched to a Vaidya exterior. This was obtained using Lie symmetry methods \cite{pal}.

\appendix

\section{Solutions for Model A}\label{sec:A}

\subsection{Static solution $X_{1}$}

Consider the application of the Lie point symmetry vector $X_{1}$ which leads to the the static solution $B\left(  t,r\right)  = B\left(  r\right)  .$
Equation (\ref{sw.09}) becomes

\begin{equation}
BB_{r}\left(  BB_{r}+2r\left(  B_{r}\right)  ^{2}-rBB_{rr}\right)  = 0,
\end{equation}

which provides the solutions
\[
B_{1}\left(  r\right)  = \lambda_{0}~ ,~~ B_{2}\left(  r\right) = -\frac{1}{\lambda_{1}r^{2}+\lambda_{0}}.
\]

\subsection{Similarity solution of $X_{1}+\alpha X_{3}$}

The application of the Lie point symmetry vector $X_{1}+\alpha X_{3}$ gives
the exact solution%
\begin{equation}
B\left(  t,r\right)  =\left(  \lambda_{1}r^{2}+\lambda_{0}-\alpha t\right)
^{-1}.
\end{equation}

\subsection{Similarity solution of $X_{1}+\alpha X_{4}$}

From the Lie point symmetry $X_{1}+\alpha X_{4}$ we find the exact solutions
\begin{equation}
B\left(  t,r\right)  =\left(  4\alpha^{2}\lambda_{1}\exp\left(  -\frac
{r^{2}-2\alpha t}{4\alpha^{2}}\right)  +\lambda_{0}\right)  ^{-1}.
\end{equation}

\subsection{Scaling solution $X_{7}$}

Reduction with the Lie symmetry vector $X_{7}$ leads to the scaling solution

\begin{eqnarray}
B\left( t,r\right)  &=& -\bigg( \lambda_{1}\int\frac{\exp\left(  \left(8u\right)  ^{-1}\right)  }{u^{\frac{3}{4}}}\times \nonumber \\
&& \exp\left(-\frac{1}{4}\int\frac{\sqrt{17u^{4}+14u^{2}+1}}{u^{3}}\right)  +\lambda_{0}\bigg)^{-1},
\end{eqnarray}

where $u = r/t $.

\subsection{Similarity solution of $X_{1}+\alpha X_{4}+\beta X_{3}$}

The application of the Lie point symmetry $X_{4}+\beta X_{2}$ provides the
similarity solutions

\begin{equation}
B\left(  t,r\right)  =\frac{\exp\left(  \frac{\sqrt{2}t}{2\sqrt{\alpha}}%
+\frac{r^{2}}{2\alpha}\right)  }{\sqrt{\left(  \lambda_{1}\exp\left(
\frac{\sqrt{2}t}{\sqrt{\alpha}}\right)  -\lambda_{2}\right)  }}.
\end{equation}

\subsection{Similarity solution of $X_{1}+\alpha X_{5}$}

The exact solution which follows from the application of the symmetry vector
$X_{1}+\alpha X_{5}$ is

\begin{equation}
B\left(  t,r\right)  =\frac{2}{\alpha}\left(  r^{2}-t^{2}\right)  ^{-1}.
\end{equation}

\subsection{Similarity solution of $X_{1}+\alpha X_{5}+\beta X_{6}$}

Hence, the exact solution which follows from the application of the symmetry
vector $X_{1}+\alpha X_{5}+\beta X_{6}$ is expressed as follows,

\begin{eqnarray}
B\left(  t,r\right)  &=& 6\bigg( 2\beta t^{3}-6\beta r^{2}t-3\alpha t^{2} \nonumber \\
&& +\frac{\left(  \alpha^{2}-4r^{2}\beta^{2}-8\beta^{2}\lambda_{1}\right)
^{\frac{3}{2}}}{12\beta^{2}}+\alpha\frac{r^{2}}{2}+\lambda_{0}\bigg)^{-1}.
\end{eqnarray}

\subsection{Similarity solution of $X_{4}+\alpha X_{5}+\beta X_{6}$}

From the application of the Lie symmetry vector $X_{4}+\alpha X_{5}+\beta
X_{6}$ it follows,

\begin{eqnarray}
B\left(t, r\right) &=& 6\bigg( 2\beta t^{2}r^{2}-\beta r^{4}-2\alpha tr^{2}-\frac{2}{3}\alpha t^{3} \nonumber \\
&& +\frac{\beta}{3}t^{4}+\frac{\alpha^{2}}{\beta}t^{2}+\lambda_{1}\xi+\lambda_{0}\bigg) ^{-1}.
\end{eqnarray}

\section{Solutions for Model B}\label{sec:B}

\subsection{Static solution $Y_{1}$}

The similarity solution which follows from the application of the Lie symmetry
vector $Y_{1}$ is static and it is expressed as follows

\[
B\left(t, r\right)  =\frac{\lambda_{0}}{1+\omega^{2}r^{2}}~,~or,~
B\left(t, r\right)  =\frac{1}{\lambda_{1}r^{2}+\lambda_{0}}
\]

\subsection{Similarity solution $Y_{1}+\alpha Y_{6}$}

Application of the Lie symmetry vector $Y_{1}+\alpha Y_{6}$ gives the
similarity solution

\begin{eqnarray}
B\left(  t,r\right) &=& -\frac{8\alpha\omega^{2}}{\omega^{2}r^{2}+1}\left(\lambda_{1}e^{-4z\alpha\omega^{2}}-\lambda_{0}\right)  ^{-1}~,~ or \nonumber \\
B\left(  t,r\right) &=& -\frac{2\left(  4\omega^{2}\alpha+1\right)  }{\omega^{2}r^{2}+1}\left(  \lambda_{1}e^{-\frac{\left(  4\omega^{2}\alpha+1\right)  }{2a}z}-\lambda_{0}\right)  ^{-1},
\end{eqnarray}
where $z=-\frac{1}{4\alpha\omega^{2}}\left(  \ln\left(  \frac{\omega^{2}r^{2}-1}{\omega^{2}r^{2}+1}\right)  +4\alpha\omega^{2}t\right)  $.

\subsection{Similarity solution $Y_{2} + \alpha Y_{6}$}

Reduction with the Lie symmetry vector $Y_{2} + \alpha Y_{6}$ provides the
similarity solution

\begin{equation}
B\left(t, r\right)  =\left(  \omega^{2}r^{2}-1\right)  ^{-\frac{1}{4\alpha\omega^{2}}}\left(  \omega^{2}r^{2}+1\right)  ^{\frac{1}{4\alpha
\omega^{2}}-1}\bar{B}\left(  t\right) ,
\end{equation}

where

\begin{eqnarray}
\bar{B}\left(  t\right)  &=& \bigg( \lambda_{1}\sin{\left(\sqrt{\frac{2\left(2\alpha\omega^{2}-1\right)}{\alpha}}t\right)}  \nonumber \\
&& - \lambda_{2}\cos{\left(\sqrt{\frac{2\left( 2\alpha\omega^{2}-1\right)  }{\alpha}}t\right)}\bigg)^{\frac{1}{2\left(  2\alpha\omega^{2}-1\right)  }}
\end{eqnarray}

\subsection{Similarity solution $Y_{1} + \alpha Y_{2} + \beta Y_{6}$}

Finally, reduction with respect to the Lie invariants of the vector field $Y_{1} + \alpha Y_{2} + \beta Y_{6}$ gives the similarity solution

\begin{equation}
B\left(t, r\right) = \lambda_{0}\frac{\left(  \omega^{2}r^{2}+1\right)
^{\frac{\alpha}{4\beta\omega^{2}}-1}}{\left(  \omega^{2}r^{2}-1\right)^{\frac{\alpha}{4\beta\omega^{2}}}}e^{-\left(  \alpha-4\beta\omega^{2}\right)  \zeta}
\end{equation}

where $\zeta=\frac{1}{4\beta\omega^{2}}\left(  \ln\left(  \frac{\omega^{2}r^{2}-1}{\omega^{2}r^{2}+1}\right)  +4\beta\omega^{2}t\right) $.

\section{Solutions for Model C}\label{sec:C}

\subsection{Similarity solutions $Z_{2}$}

Reduction with respect to the symmetry vector leads to the similarity solution

\begin{equation}
B\left(  t,r\right)  = r^{-\frac{1}{N+1}}\left(  \lambda_{1}\left(t-t_{0}\right)  \right)  ^{\frac{2N+1}{4N^{2}+3N+1}}.
\end{equation}

\section*{Acknowledgements}

RB and MG acknowledge support from the office of the Deputy Vice-Chancellor for Research and Innovation at the Durban University of Technology. \\

\section*{Data Availability Statement}

This manuscript has no associated data or the data will not be deposited. (Authors' comment: All data was obtained using the formulae explicitly given in the article.) \\



\newpage

\begin{thebibliography}{99}

\bibitem{opp} J. R. Oppenheimer, H. Snyder, Phys. Rev. {\bf 56}, 455 (1939)

\bibitem{vaid} P. C. Vaidya, Proc. Indian Acad. Sci. A {\bf 33}, 264 (1951)

\bibitem{bon} W. B. Bonnor, A. K. G. de Oliveira, N. O. Santos, Phys. Rep. {\bf 181}, 269 (1989)

\bibitem{pret} J. M. Z. Pretel, M. F. A. da Silva, Mon. Not. R. Astron. Soc. {\bf 495}, 5027 (2020)

\bibitem{bog} R. S. Bogadi, M. Govender, S. Moyo, Eur. Phys. J. C {\bf 81}, 922 (2021)

\bibitem{glas} E. N. Glass, Phys. Lett. A {\bf 867}, 351 (1981)

\bibitem{san} N. O. Santos, Mon. Not. R. Astron. Soc. {\bf 216}, 403 (1985)

\bibitem{nai} N. F. Naidu, M. Govender, K. S. Govinder, Int. J. Mod. Phys. D {\bf 15}, 1053 (2006)

\bibitem{thir1} S. Thirukkanesh, S.S. Rajah, S.D. Maharaj, J. Math. Phys. {\bf 53}, 032506 (2012)

\bibitem{mah1} S. D. Maharaj, M. Govender, Pramana J. Phys. {\bf 54}, 715 (2000)

\bibitem{thir2} S. Thirukkanesh, S. Moopanar, M. Govender, Pramana J. Phys. {\bf 79}, 223 (2012)

\bibitem{her1} L. Herrera, G. Le Denmat, N. O. Santos, Phys. Rev. D {\bf 79}, 087505 (2009)

\bibitem{her2} L. Herrera, A. Di Prisco, N. O. Santos, Gen. Relativ. Grav. {\bf 42}, 1585 (2010)

\bibitem{her3} L. Herrera, G. Le Denmat, N.O. Santos, Gen. Relativ. Grav. {\bf 44}, 1143 (2012)

\bibitem{mah2} S. D. Maharaj, M. Govender, Int. J. Mod. Phys. D {\bf 14}, 667 (2005)

\bibitem{gov} M. Govender, R. S. Bogadi, D. B. Lortan, S. D. Maharaj, Int. J. Mod. Phys. D {\bf 25}, 1650037 (2016)

\bibitem{red} K. P. Reddy, M. Govender, W. Govender, S. D. Maharaj, Annals Phys. {\bf 429}, 168458 (2021)

\bibitem{mso} A. M. Msomi, K. S. Govinder, S. D. Maharaj, Int. J. Theor. Phys. {\bf 51}, 1290 (2012)

\bibitem{abe1} G. Z. Abebe, K. S. Govinder, S. D. Maharaj, Int. J. Theor. Phys. {\bf 52}, 3244 (2013)

\bibitem{abe2} G. Z. Abebe, S. D. Maharaj, K. S. Govinder, Gen. Relativ. Grav. {\bf 46}, 1650 (2014)

\bibitem{mau1} S. K. Maurya, Y. K. Gupta, S. Ray, D. Deb, Eur. Phys. J. C {\bf 76}, 693 (2016) 

\bibitem{kar} K. R. Karmarkar, Proc. Indian Acad. Sci. A {\bf 27}, 56 (1948)

\bibitem{pan} S. N. Pandey, S. P. Sharma, Gen. Relativ. Gravit. {\bf 14}, 113 (1981)

\bibitem {bha} P. Bhar, Ksh. Newton Singh, F. Rahaman, N. Pant, S. Banerjee, Int. J. Mod. Phys. D. {\bf 26}, 1750078 (2017)

\bibitem{mau2} S. K. Maurya, S. D. Maharaj, Eur. Phys. J. C {\bf 77}, 328 (2017)

\bibitem{abe3} G. Z. Abebe, S. D. Maharaj, K. S. Govinder, Eur. Phys. J. C {\bf 75}, 496 (2015)  

\bibitem{abe4} G. Z. Abebe, S. D. Maharaj, Eur. Phys. J. C {\bf 79}, 849 (2019) 

\bibitem{GG} K. S. Govinder, M. Govender, Gen. Relativ. Gravit. {\bf 44}, 174 (2012)

\bibitem{ch5} A. Paliathanasis and M. Tsamparlis, IJGMMP \textbf{11,} 1450037 (2014)

\bibitem{ch6} T. Christodoulakis, N. Dimakis, P.A. Terzis, G. Doulis, Th. Grammenos, E. Melas and A.\ Spanou, J.\ Geom. Phys. \textbf{71,} 127 (2013)

\bibitem{ch7} B. Muatjetjeja and C.M. Khalique, Acta Mathematica Scientia \textbf{32,} 1959 (2012)

\bibitem{ch8} E.D. Fackerell and D. Hartley, Gen. Relativ. Gravit. \textbf{32,} 857 (2000)

\bibitem{ch9} A. Paliathanasis, Mod. Phys. Lett. A {\bf 37}, 2250119 (2022)

\bibitem{ch10} M. Tsamparlis and A. Paliathanasis, Symmetry \textbf{10,} 233 (2018)

\bibitem{ch11} A. Janda, Acta Physica Polonica B \textbf{38,} 3961 (2007)

\bibitem{Stephani} H. Stephani, Differential Equations: Their Solutions Using Symmetry, Cambridge University Press, New York, (1989)

\bibitem{Bluman} G. W. Bluman and S. Kumei, Symmetries of Differential Equations, Springer-Verlag, New York, (1989)

\bibitem{ibra} N. H. Ibragimov, CRC Handbook of Lie Group Analysis of Differential Equations, Volume I: Symmetries, Exact Solutions, and
Conservation Laws, CRS Press LLC, Florida (2000)

\bibitem{olver} P. J. Olver, Applications of Lie Groups to Differential Equations, Springer-Verlag, New York, (1993)
 
\bibitem{nai2} N. F. Naidu, M. Govender, S. D. Maharaj, Eur. Phys. J. C {\bf 78}, 48 (2018)

\bibitem{pal} A. Paliathanasis, M. Govender, G. Leon, Eur. Phys. J. C {\bf 81}, 718 (2021) 

\end{thebibliography}
\end{document}